\documentstyle[12pt,bezier]{article}
\begin{document}
\def \inbar{\vrule height1.5ex width.4pt depth0pt}
\def \xC{\relax\hbox{\kern.25em$\inbar\kern-.3em{\rm C}$}}
\def \xR{\relax{\rm I\kern-.18em R}}
\newcommand{\xZ}{Z \hspace{-.08in}Z}
\newcommand{\xbe}{\begin{equation}}
\newcommand{\xee}{\end{equation}}
\newcommand{\xbea}{\begin{eqnarray}}
\newcommand{\xeea}{\end{eqnarray}}
\newcommand{\xnn}{\nonumber}
\newcommand{\xkt}{\rangle}
\newcommand{\xbr}{\langle}
\newcommand{\xlll}{\left( }
\newcommand{\xrrr}{\right)}
\newcommand{\xcun}{\mbox{\footnotesize${\cal N}$}}
\newcommand{\cun}{\mbox{\footnotesize${\cal N}$}}
\title{Comment on Identical Motion in Classical and Quantum Mechanics}
\author{Ali Mostafazadeh\thanks{E-mail address: 
amostafazadeh@ku.edu.tr}\\ \\
Department of Mathematics, Ko\c{c} University,\\
Istinye 80860, Istanbul, TURKEY}
\date{ }
\maketitle

\begin{abstract}
Makowski and Konkel [Phys.\ Rev.\ A {\bf 58}, 4975 (1998)] 
have obtained certain classes of potentials which lead to identical
classical and quantum Hamilton-Jacobi equations. We obtain the most
general form of these potential.
\end{abstract}
PACS numbers: 03.65.Sq, 03.65.Bz


In their recent paper \cite{m-k}, Makowski and Konkel study the
class of potentials allowing for a quantum potential $Q$ which only
depends on time,
	\xbe
	Q=K(t)\;.
	\label{Q=const}
	\xee
Since the quantum effects are related to the quantum force $-\nabla Q$,
the corresponding systems have identical classical and quantum dynamics.
In the polar representation where the wave function takes the form $\Psi(\vec x,t)=
R(\vec x,t)\exp[i\varphi(\vec x,t)/\hbar]$ the Schr\"odinger equation is
written as
	\xbea
	\frac{\partial R^2}{\partial t}+\nabla\cdot\left(R^2\nabla\frac{\varphi}{m}
	\right)&=&0\;,
	\label{2a}\\
	\frac{\partial\varphi}{\partial t}+\frac{(\nabla\varphi)^2}{2m}+V+Q&=&0\;.
	\label{2b}
	\xeea
Here $R$ and $\varphi$ are real-valued functions and the quantum potential $Q$ 
is defined by $Q(\vec x,t):=-\hbar^2\nabla^2 R/(2m R)$. 

The classification of all the potentials with the above mentioned property 
is equivalent to finding the general solution of Eqs.~(\ref{Q=const}),~(\ref{2a})
and (\ref{2b}) for the three real unknowns, $R,~\varphi,$ and $V$. The complete
classification for the case $K=0$ is given in Ref.~\cite{npb} which predates
the work of Makowski and Konkel \cite{m-k}. In Ref.~\cite{npb} the analogous 
problem for Klein-Gordon equation is also addressed. Moreover a new semiclassical 
perturbation theory around these potentials is outlined, and its application to 
quantum cosmology is discussed. 

The results of Makowski and Konkel \cite{m-k} can also be generalized to give 
the complete classification of potentials which allow for identical classical
and quantum dynamics in arbitrary dimensions. Following Ref.~\cite{npb}, we shall
refer to these potential as {\em semiclassical potentials}. 

As explained by Makowski and Konkel \cite{m-k}, for the case of stationary
states, where $\partial R/\partial t=0$ and $\varphi(\vec x,t)=-Et+S(\vec x)$,
Eq.~(\ref{2a}) reduces to
	\xbe
	\nabla\cdot(R^2\nabla S)=0\;.
	\label{3}
	\xee
Makowski and Konkel \cite{m-k} obtain a class of solutions of this
equation by solving 
	\xbe
	R^2\nabla S={\rm const}\;.
	\label{5}
	\xee
Therefore, they restrict their analysis to a set of particular solutions of
Eq.~(\ref{3}). This is, however, not necessary. The general solution of 
Eq.~(\ref{3}) can be easily obtained by making the following change of 
dependent variable:
	\xbe
	S\to \tilde S:=RS\;.
	\label{tilde-s}
	\xee
Let us first introduce
	\[\lambda:=\frac{\sqrt{2mK}}{\hbar}\;,\]
in terms of which Eq.~(\ref{Q=const}) takes the form
	\xbe
	\nabla^2 R+\lambda^2 R=0\;.
	\label{r-eq}
	\xee
Now substituting $S=\tilde S/R$ in (\ref{3}) and making use of Eq.~(\ref{r-eq}),
we obtain
	\xbe
	\nabla^2 \tilde S+\lambda^2 \tilde S=0\;.
	\label{tilde-s-eq}
	\xee
Therefore in view of Eqs.~(\ref{2b}), (\ref{tilde-s}), and (\ref{r-eq}),
the most general potential allowing for identical classical and quantum
dynamics for a stationary state is given by
	\xbe
	V=E-\frac{1}{2m}\left\{(\hbar\lambda)^2+
	\left[\nabla\left(\frac{\tilde S}{R}\right)\right]^2\right\}\;,
	\label{6'}
	\xee
where $R$ and $\tilde S$ are solutions of Eqs.~(\ref{r-eq}) and (\ref{tilde-s-eq}),
respectively. Note that Eq.~(\ref{6'}) is valid in any number of dimensions.

More generally, we can use the analog of the change of variable (\ref{tilde-s}), namely
	\xbe
	\varphi\to\tilde\varphi:=R\varphi\;,
	\label{tilde-phi}
	\xee
to handle the general problem, where the wave function $\Psi$ does not represent
a stationary state. In this case, Eq.~(\ref{r-eq}) still holds, but $\lambda$ is 
a function of time. Using (\ref{tilde-phi}) and (\ref{r-eq}), Eqs.~(\ref{2a}) 
and (\ref{2b}) take the form
	\xbea
	&&\nabla^2 \tilde\varphi+\lambda^2 \tilde\varphi=-2m
	\frac{\partial R}{\partial t}\;,
	\label{2a'}\\
	&& V=-\frac{\partial}{\partial t}\left(\frac{\tilde\varphi}{R}\right)
	-\frac{1}{2m}\left\{(\hbar\lambda)^2+ 
	\left[\nabla\left(\frac{\tilde\varphi}{R}\right)\right]^2\right\}
	\;.
	\label{2b'}
	\xeea
Therefore the set of all the semiclassical potentials are classified by the
solutions of Eqs.~(\ref{r-eq}) and (\ref{2a'}). Note that these equations are not
evolution equations. Eq.~(\ref{r-eq})  may be viewed as a constraint equation in which
$t$ enters as a parameter through the dependence of $\lambda$ on $t$. Once the boundary
conditions of this equation are chosen, it can be solved using the known methods of
solving linear partial differential equations with `constant' coefficients. The solution
is then used to evaluate the right hand side of Eq.~(\ref{2a'}). The latter is a 
nonhomogeneous linear partial differential equation. It can be solved using the
well-known Green's function methods. Again it is the boundary conditions that
determine the solution. 

The above analysis shows that it is the choice of the function $K$ or alternatively
$\lambda$ together with the boundary conditions of Eqs.~(\ref{r-eq}) and (\ref{2a'})
that determine the set of potentials which allow for identical classical and 
quantum dynamics.

We wish to conclude this article by the following remarks.
	\begin{itemize}
	\item[1.)] Makowski and Konkel conclude their paper \cite{m-k} emphasizing 
that ``{\em A number of additional potentials would be found if new solutions of 
Eq.~(2a)} [This is our Eq.~(2)] {\em were obtained. This, however, can be a difficult
task.}'' We have shown that a simple change of the dependent variable $\varphi$, namely
$\varphi\to\tilde\varphi:=R\varphi$, eases this `difficulty' and leads to a complete 
classification of all such potentials.
	\item[2)] Following the above remark, Makowski and Konkel write: ``{\em
Among the potentials derived here we have not found any example from the known set
of potentials implying bound states, e.g., Coulomb, Morse, or P\"oschl-teller. This
likely follows from the fact that most stationary states of physical interest have no
classical limit.}'' In this connection, we must emphasize that for the semiclassical 
potentials leading to identical classical and quantum dynamics, the amplitude $R:=
|\Psi|$ of the wave function of a stationary state satisfies Eq.~(7). This
is just the eigenvalue equation for the Laplacian in $\xR^n$. It is well-known that
this equation does not admit a  solution corresponding to a bound state.
	\item[3)] It is well-known that shifting the Hamiltonian $H$ by a 
time-dependent multiple of the identity operator, i.e.,
	\xbe
	H\to H'=H+f(t)\;.
	\label{trans-h}
	\xee
leaves all the physical quantities of the system invariant \cite{bohm-qm}. This is in
fact true both in quantum and classical mechanics. In quantum mechanics, such a 
transformation corresponds to a phase transformation 
	\xbe
	\Psi\to\Psi'= e^{i\zeta(t)}\Psi
	\label{trans}
	\xee
of the Hilbert space, where $\zeta(t):=-\int_0^t f(t')dt'$. Therefore, it leaves the 
quantum states, the expectation values of the observables, and the excitation energies
invariant. It also leaves the quantum potential invariant. But it does change the 
classical potential according to
	\xbe
	V\to V'=V+f(t)\;.
	\label{trans-v}
	\xee
In fact, the effect of such a transformation on Eqs.~(2) and (3) is the shift 
(\ref{trans-v}) of the potential.

Now consider the case that the quantum potential is a function of time, $Q=K(t)$.
Then we can make a phase transformation (\ref{trans}) of the Hilbert space with 
$\zeta(t)=\int_0^t K(t')dt'$ in (\ref{trans}), so that $f=-K$ and the total 
potential $V+Q=V+K$ in Eq.~(3) is transformed to $V'+Q-K=V$. Therefore, the dynamics
of a state with time-dependent quantum potential and a state with a zero quantum 
potential are equivalent. In general, we can make a phase transformation of the 
Hilbert space which effectively removes such a quantum potential. Therefore,
as far as the physical quantities are concerned the case $Q=K(t)$
is equivalent to the case $Q=0$. The latter has been thoroughly studied in 
Ref.~\cite{npb}.
	\end{itemize}


\begin{thebibliography}{9}
\bibitem{m-k} A.\ J.\ Makowski and S.\ Konkel, Phys.\ Rev.\ A {\bf 58}, 4975 (1998).
\bibitem{npb} A.~Mostafazadeh, Nucl.\ Phys.\ {\bf B 509}, 529-555 (1998). Note that this
article was submitted for publication in January 1997 and published in February 1998,
whereas Ref.~\cite{m-k} was submitted in May 1998 and published in December 1998.
\bibitem{bohm-qm} A.~Bohm, {\em Quantum Mechanics: Foundations and Applications,}
third edition (Springer-Verlag, Berlin, 1993)
\end{thebibliography}
\end{document}